\documentclass{appolb}

\usepackage{graphicx}
\usepackage{amsmath}
\usepackage{cite}
\usepackage{hyperref}

\usepackage{float}

\title{\MakeUppercase{Measurement of the mass of Higgs boson through HZ production at FCC-ee}}

\author{I.~KAHRAMAN$^{a,}$ 
\thanks{isinsukahraman@gmail.com}
and 
O.~\c{C}AKIR$^{b,}$
\thanks{ocakir@science.ankara.edu.tr}
\address{$^a$Graduate School of Natural and Applied Sciences, Ankara University, Ankara, Turkiye}
\address{$^b$Department of Physics, Ankara University, Ankara, Turkiye}}

\begin{document}

\maketitle

\begin{abstract}
The associated production of a Higgs boson with a $Z$ boson in the collisions of electron and positron beams at Future Circular Collider (FCC-ee) have been studied. Here, the $Z$ boson decays into dileptons, while the Higgs boson decays to all possible channels, mostly to the $b\bar{b}$. The collisions provide a clean and powerful channel to probe the ($HZZ$) coupling at future lepton colliders. Following the event generation, the analysis is performed using the recoil mass method, which allows model-independent reconstruction of the Higgs boson depending on the kinematics of the final-state leptons. This method enables precise identification of the Higgs signal peak independent of its decay mode and significantly reduces systematic uncertainties. The recoil mass distributions from the signal process ($e^+e^- \to HZ$, $Z \to l^+l^-$) and the main backgrounds ($ZZ$, $WW$, $t\bar{t}$ and other standard model processes) have been analyzed using a dedicated analysis code. Monte Carlo simulations corresponding to an integrated luminosity of $5$ ab$^{-1}$ have been used for the analysis, assuming the high performance of the IDEA detector concept. The results are presented for center-of-mass energies of $\sqrt{s} = 240$ GeV and $\sqrt{s} = 365$ GeV to compare the sensitivities and highlight the potential of future $e^+e^-$ colliders in probing the $HZZ$ interaction with high precision.
\end{abstract}

\section{Introduction}

The Higgs boson \cite{Higgs1964} is considered to be sensitive to new physics beyond the standard model (BSM) of particle physics. This particle was discovered by the ATLAS and CMS experiments \cite{ATLAS_CMS} at the Large Hadron Collider (LHC). Higgs physics is related to numerous questions, then searches of its properties at future colliders may provide some answers.

The measurement of single Higgs boson production cross section, using recoil mass method, set the scale for Higgs coupling measurements. This method enables the study of the Higgs boson without the need to reconstruct its decay products and it plays a central role in determining the Higgs mass and measuring the inclusive cross section of the process, which in turn provides direct access to the $HZZ$ coupling. 

The future circular $e^+e^-$ collider (FCC-ee) \cite{Abada2019a,Abada2019b} would allow the Higgs scalar sector to be probed with high precision beyond that achievable at the LHC. Focus on the Higgs searches at the $e^+e^-$ collider would be model independent determination of the Higgs couplings to gauge bosons and fermions.

In this study, the aim is to evaluate the performance of the recoil mass method for measuring the Higgs mass and cross section of the $HZ$ production at two proposed operating energies of future circular electron-positron collider (FCC-ee) with $240$ GeV and $365$ GeV stages. The recoil mass distributions from the signal $(e^+ e^- \to HZ, Z \to l^+ l^-)$ and the backgrounds (mainly $ZZ$, $WW$, $t\bar{t}$  and others) have been analyzed and the parameters have been optimized to obtain more precise results.  Using simulated signal and background events, we analyze how the measurement sensitivities vary with center-of-mass energies. This method plays an important role in probing the $HZZ$ interactions. 

\section{Associated production and decay process}

The Standard Model (SM) of particle physics describes three fundamental interactions between elementary particles. It is a renormalizable gauge theory based on the symmetry group $SU(3)_C\times SU(2)_L\times U(1)_Y$, see Ref. \cite{PDG2024}. The Higgs scalar sector of the model is central for the physics program of present and future colliders.

The associated production of the Higgs boson with a $Z$ boson in electron-positron collider is described at leading order by the Feynman diagram involving s-channel $Z$ boson exchange. The cross section for this process is sensitive to the strength of the $HZZ$ coupling, which can be parameterized in the Lagrangian as:
$$
\mathcal{L}_{HZZ} = -g_{HZZ} H Z_\mu Z^\mu  \eqno{(1)}
$$
where $g_{HZZ}$ is the coupling constant that determines the interaction strength between the Higgs and $Z$ bosons. In the production mode the weak neutral current interactions related to the charged leptons are described by Lagrangian:
$$
\mathcal{L}_{Zll} = -\bar{l}\gamma^\mu(g_{L} P_L +g_R P_R) l Z_\mu  \eqno{(2)}
$$
where $P_{L,R}=(1\mp\gamma_5)/2$ are left and right chiral projections and  $g_{L,R}$ are the chiral couplings of the $Z$ boson to corresponding leptons. One may use the notation for these couplings such as $g_{L,R}=g/\cos\theta_W (T_{3\,L,R}+\sin^2\theta_W)$, where $T_3$ is the third component of weak isospin of the corresponding lepton, and $\theta_W$ is the weak mixing Weinberg angle.

In the decay channel, the interaction of the Higgs boson with fermions is described by the Yukawa coupling term in the Lagrangian:
$$
\mathcal{L}_{Hff} = -\sum_f \frac{m_f}{v} H \bar{f} f    \eqno{(3)}
$$
where $ m_f $ is the mass of the fermion ($f$), $v$ is the vacuum expectation value of the Higgs field ($H$). This term implies that the coupling strength of the Higgs boson to fermions is proportional to the fermion mass. As a result, the dominant decay channel of the SM Higgs boson is into a pair of bottom quarks ($H \to b\bar{b}$), with a branching ratio of approximately 58\%, see Ref. \cite{PDG2024}. Other important decay modes include $H \to WW^*$, $gg$, $\tau^+\tau^-$, $ZZ^*$ and others.

The cross section for the associated $HZ$ production is proportional to the square of the coupling:
$$
\sigma(e^+e^- \rightarrow ZH) \propto |g_{HZZ}|^2
\eqno{(4)} $$  
which highlights that any measurement of the production rate directly constrains the $HZZ$ coupling.
The cross section for the exclusive fermionic or bosonic final ($F$) state decays ($H\to FF$) 
can be expressed as
$$
\sigma(e^+e^- \rightarrow ZH) \times BR(H\to FF)\propto |g_{HZZ}|^2 |g_{HFF}|^2/\Gamma_H
  \eqno{(5)}  $$

At center-of-mass energy of 240 GeV, the cross section of the process is maximized, and the recoil mass resolution become optimal due to the limited phase space for the final states. At a higher energy of 365 GeV, the process benefits from higher recoil energies and extended kinematic reach, though the cross section is relatively smaller. These features make both energy points complementary in probing different aspects of Higgs physics.

\section{Recoil mass method}

We focus on a model independent approach to Higgs boson measurement using the recoil mass method. This allows for the identification of Higgs events by reconstructing only the decay products of the associated $Z$ boson, particularly in the leptonic decay mode ($Z \to l^+l^-$), without depending on the specific Higgs decay channel.
This approach enables precise determination of the inclusive Higgs production cross section $\sigma(e^+e^- \to HZ)$ and provides direct sensitivity to the $HZZ$ coupling strength $g_{HZZ}$. By analyzing the recoil mass spectrum, one can extract both the production rate and test the structure of the $HZZ$ interaction with minimal model dependence. This is particularly advantageous at lepton colliders such as FCC-ee, where clean experimental environments allow high-precision measurements, see Ref. \cite{Azzurri2022}.

The recoil mass squared expression is defined as:
$$
M^2_{\text{recoil}} = (\sqrt{s} - E_{ll})^2 - |\vec{p}_{ll}|^2=s-2\sqrt{s}E_{ll}+m_{ll}^2
\eqno{(6)}
$$
where $E_{ll}$ and $\vec{p}_{ll}$ are the energy and momentum of the lepton pair originating from the $Z$ boson decay, $m_{ll}$ is the invariant mass of two leptons showing peak around $m_Z$, and $\sqrt{s}$ is the center-of-mass energy of the collision. This recoil mass method allows the reconstruction of the Higgs boson mass peak without requiring direct reconstruction of its decay products, enabling a model independent measurement.

The measurement of the single Higgs boson production cross section using recoil mass method determines the scale of Higgs coupling measurements. The $HZ$ events can be determined easily by counting the normalized entries from the collection around Higgs mass. The number of events allows the determination of $HZ$ cross section $\sigma(HZ)$. This cross section is proportional to the square of the coupling $g_{HZZ}$. In most studies, the recoil mass method has been applied to the process $e^+e^-\to HZ$, where the $Z$ boson decays into two oppositely charged leptons, see Ref. \cite{Abada2019a} and \cite{Ruan2014}. However, a detailed study  have shown that hadronic decays of the $Z$ boson enhance sensitivity due to their larger branching ratio, see Ref.  \cite{Thomson2016}.

\section{Associated Higgs production}

\begin{figure}[htb]
\centerline{
\includegraphics[width=0.45\linewidth]{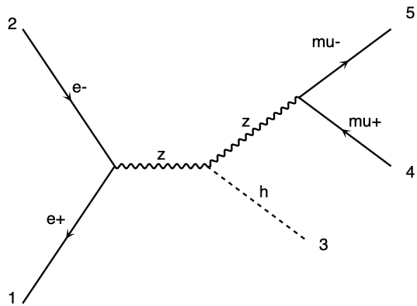}}
\caption{Relevant diagrams including HZZ (or hzz) interaction vertex and leading to $h\mu^+\mu^-$ state, where $\mu^\pm$ is denoted as mu+/mu- in the Figure due to particle definition in event generator.}
\label{fig:feyn1}
\end{figure}

\begin{figure}[htb]
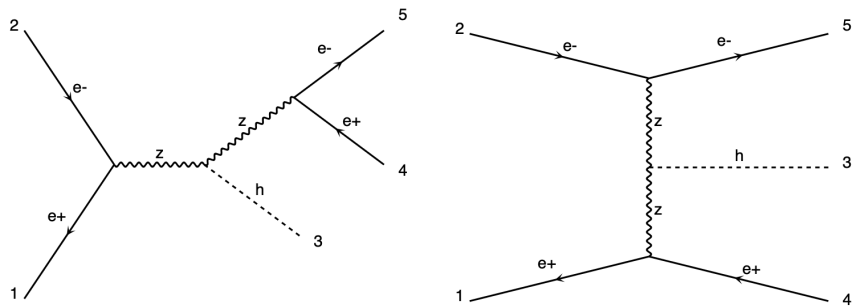

\centerline{
\includegraphics[width=0.45\linewidth]{feyn_hz2ee.pdf}
\includegraphics[width=0.45\linewidth]{feyn_hee.pdf}}
\caption{Relevant diagrams including HZZ (or hzz) interaction vertex and leading to $he^+e^-$ state, where $e^\pm$ is denoted as e+/e- in the Figure due to particle definition in event generator.}
\label{fig:feyn2}
\end{figure}

In $e^+ e^-$ collisions at $240$ GeV and $365$ GeV, the interaction vertex $HZZ$ can be investigated through two main production mechanisms which corresponds to $ZH$ production (Higgsstrahlung) (as shown in Fig. \ref{fig:feyn1} and left panel of Fig. \ref{fig:feyn2} ) and the $ZZ$ fusion process shown in the right panel of Fig. \ref{fig:feyn2}. The cross section for the s-channel $ZH$ production process is higher than t-channel $ZZ$ fusion process as indicated in Table \ref{tab:cross_sections_ISR}. 
The cross sections have been calculated separately for the di-electron and di-muon channel, and included both s-channel and t-channel contributions in di-electron channel.

We have calculated cross sections of single Higgs boson production processes at $\sqrt{s} = 240$ GeV and $\sqrt{s} = 365$ GeV with and without initial state radiation (ISR) and beamstrahlung (BS) effects. We find the ratio of cross sections $\sigma(\mathrm{ISR+BS})/\sigma$ to be around $0.85$ at 
$\sqrt{s}=240$ GeV, and around $1.04$ at $365$ GeV. In these calculations we have used MadGraph5 \cite{Alwall2011} with the inclusion of ISR+BS parametrization "fcce240ll" and "fcce365ll" as mentioned in Ref.\cite{Frixione2021}.

Table \ref{tab:cross_sections_ISR} presents the cross sections of various single Higgs boson production processes at center-of-mass energies $\sqrt{s} = 240$ GeV and $\sqrt{s} = 365$ GeV and it shows the ratio $\sigma(\mathrm{ISR+BS})/\sigma$ for these processes. The branching ratios can be used $BR(h \rightarrow b\bar{b}) = 0.58$ and $BR(Z \rightarrow l^+ l^-) = 3.37 \times 10^{-2}$ for the final state. As seen in this table, the cross section of the signal process reaches its maximum value at 240 GeV.

\begin{table}[ht]
    \centering
    \caption{Cross sections of single Higgs boson production processes at $\sqrt{s} = 240$ GeV and $\sqrt{s} = 365$ GeV. The branching ratios can be used $BR(h \to b\bar{b}) = 0.53$ and $BR(Z \to l^+l^-) = 3.37 \times 10^{-2}$ for the final state. \label{tab:cross_sections_ISR}}
    \vspace{0.5em}
    \tiny
    \begin{tabular}{|l|c|c|c|c|}
        \hline
        {Process} 
        & \multicolumn{2}{c|}{$\sqrt{s} = 240$ GeV} 
        & \multicolumn{2}{c|}{$\sqrt{s} = 365$ GeV} \\
        \cline{2-5}
        & $\sigma$ (pb) & $\sigma$(ISR+BS)/$\sigma$ 
        & $\sigma$ (pb) & $\sigma$(ISR+BS)/$\sigma$\\
        \hline
        $e^+ e^- \rightarrow hZ$ & $2.404 \times 10^{-1}$ & $0.845$ & $1.172 \times 10^{-1}$ & $1.052$ \\
        $e^+ e^- \rightarrow h e^+ e^-$ & $7.847 \times 10^{-3}$ & $0.849$ & $6.760 \times 10^{-3}$ & $0.981$ \\
        $e^+ e^- \rightarrow h \mu^+ \mu^-$ & $7.435 \times 10^{-3}$ & $0.847$ & $3.803 \times 10^{-3}$ & $1.044$ \\
        $e^+ e^- \rightarrow h \tau^+ \tau^-$ & $7.425 \times 10^{-3}$ & $0.844$ & $3.806 \times 10^{-3}$ & $1.046$ \\
        $e^+ e^- \rightarrow h \nu_e \bar{\nu}_e$ & $2.250 \times 10^{-2}$ & $0.865$ & $4.073 \times 10^{-2}$ & $0.923$ \\
        $e^+ e^- \rightarrow h \nu_{\mu,\tau} \bar{\nu}_{\mu,\tau}$ & $1.576 \times 10^{-2}$ & $0.846$ & $9.645 \times 10^{-3}$ & $1.048$ \\
        $e^+ e^- \rightarrow h( u\bar{u} / c\bar{c})$ & $2.042 \times 10^{-2}$ & $0.852$ & $1.156 \times 10^{-2}$ & $1.029$ \\
        $e^+ e^- \rightarrow h (d\bar{d} / s\bar{s})$ & $2.632 \times 10^{-2}$ & $0.847$ & $1.481 \times 10^{-2}$ & $1.032$ \\
        $e^+ e^- \rightarrow h b \bar{b}$ & $3.487 \times 10^{-2}$ & $0.847$ & $1.758 \times 10^{-2}$ & $1.046$ \\
        \hline
    \end{tabular}
\end{table}

At $\sqrt{s} = 240$ GeV, the ISR+BS effects reduce the cross sections for all processes by approximately 15\%. The ratios range between 0.842 and 0.849, indicating a consistent negative impact of ISR and BS at this center-of-mass energy. At $\sqrt{s} = 365$ GeV, the ISR+BS effects vary more significantly between processes. Some channels such as $e^+e^- \to HZ$, $H\mu^+\mu^-$, and $H\tau^+\tau^-$ experience a slight enhancement in the cross section (up to 5\%), with ratios exceeding 1.0. Conversely, processes like $h\nu_e\bar{\nu}_e$ and $hu\bar{u}/hc\bar{c}$ still exhibit a moderate reduction (around 7--9\%), as shown in Table \ref{tab:backg_cross_sections_ISR}.

This variation implies that the ISR and BS contributions depend on the specific final state and the center-of-mass energy. Hence, proper simulation and correction of ISR+BS effects are crucial in precision measurements of Higgs properties at future $e^+e^-$ colliders.

The dominant production process is clearly $e^+e^- \rightarrow HZ$, with a maximum cross section of $2.404 \times 10^{-1}$ pb at $\sqrt{s} = 240$ GeV, reducing to $1.172 \times 10^{-1}$ pb at $\sqrt{s} = 365$ GeV. This makes the 240 GeV stage optimal for model-independent Higgs studies using the recoil mass method. Other processes such as $e^+e^- \rightarrow h e^+e^-$ and $e^+e^- \rightarrow h \nu_e \bar{\nu_e}$ are representative of vector boson fusion (VBF) contributions. Their cross sections increase mildly at 365 GeV, illustrating how VBF becomes more prominent at higher energies. For example, VBF contribution to the cross section for $e^+e^- \rightarrow h \nu_e\bar\nu_e$ increases from $6.315 \times 10^{-3}$ pb at 240 GeV to $3.483 \times 10^{-2}$ pb at 365 GeV.

The inclusion of final states such as $h \rightarrow b\bar{b}$, $c\bar{c}$, $\tau^+\tau^-$, and hadronic decays of the Z boson allows for enhanced sensitivity due to higher branching ratios \cite{Thomson2016}. However, the leptonic decay channel remains the cleanest for recoil mass measurements due to its low background and excellent mass resolution.

The cross section values guide the event selection strategy and detector optimization for FCC-ee physics analyses. The complementarity between 240 GeV and 365 GeV stages highlights the benefit of running at both energy points: the lower energy for optimal recoil resolution, and the higher energy for enhanced sensitivity to alternative Higgs production mechanisms and new physics effects.

\begin{figure}[htbp]
    \centering
    \includegraphics[width=0.85\textwidth]{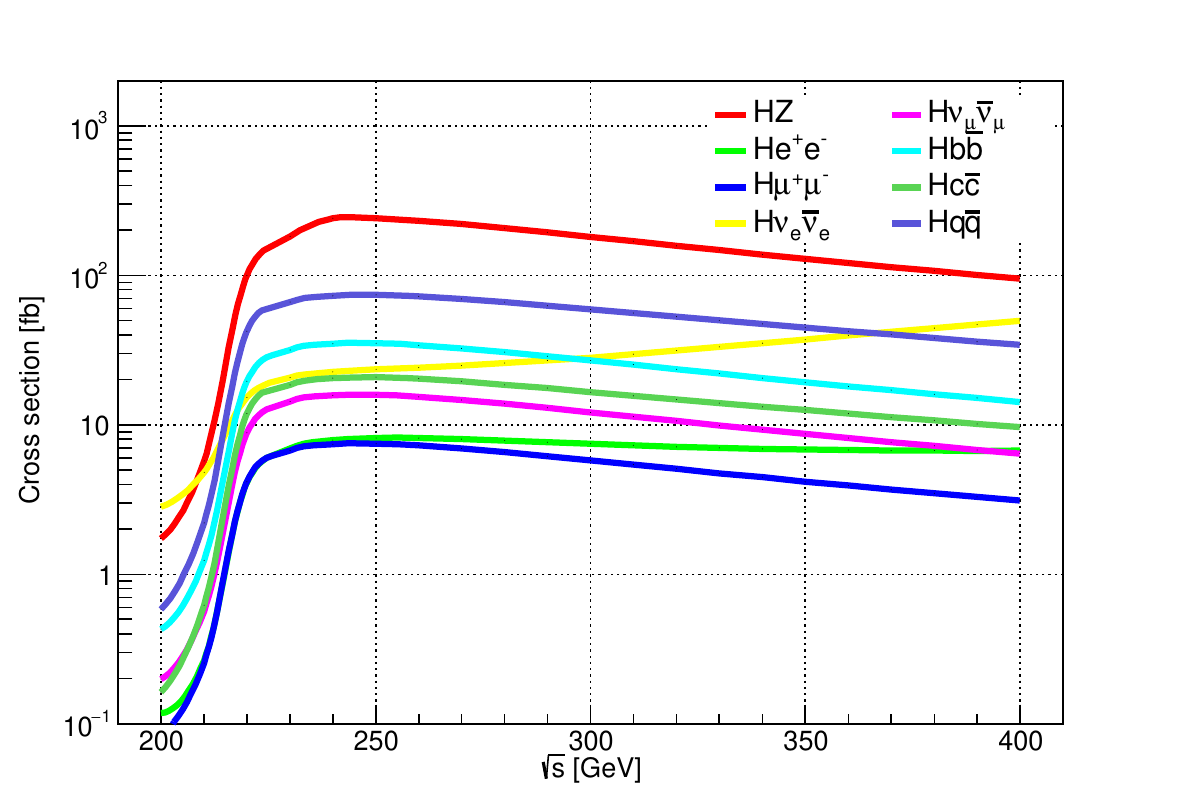}
    \caption{Cross section as a function of center-of-mass energy for various single Higgs production channels at electron-positron colliders. The highest cross section is observed for the $e^+e^- \rightarrow HZ$ process at $\sqrt{s} = 240$ GeV and 365 GeV.}
    \label{fig:singlehiggs_xs}
\end{figure}

Figure~\ref{fig:singlehiggs_xs} illustrates the variation of the cross section for single Higgs boson production as a function of center-of-mass energy. At both 240 GeV and 365 GeV energy levels, the signal process has the highest cross section compared to other single production channels. In the event generation step a minimal $p_T$ cut and $\eta$ cut have been applied on detectable particles. By default in the event generation level there is a $p_T>10$ GeV and $|\eta|<2.5$ cuts on charged leptons (electrons or muons), and applied no $p_T$ and $\eta$ cuts for bottom and charm quarks since they can also result from $W/Z/H$ hadronic decays.

The $e^+e^- \rightarrow HZ$ process (red curve) dominates the production at lower energies and reaches its maximum around $\sqrt{s} = 240$ GeV. This makes the 240 GeV stage optimal for Higgs production via the Higgsstrahlung mechanism, especially when using the recoil mass method for model-independent analyses.

Other production mechanisms, such as vector boson fusion (VBF) channels ($He^+e^-$, $H\nu_e\bar{\nu}_e$), exhibit a steady increase or flattening trend (depending on initial cuts) with rising $\sqrt{s}$ and become more significant at higher energies. This behavior is expected, as VBF processes benefit from larger phase space and higher boost at elevated energies. Consequently, the FCC-ee with 365 GeV stage provides enhanced sensitivity to these processes and is crucial for exploring $HVV$ interactions and deviations from Standard Model predictions.

Fermionic Higgs decay modes such as $H \rightarrow b\bar{b}$, $c\bar{c}$, and $\tau^+\tau^-$ remain relatively flat across the energy spectrum, with their yields primarily governed by branching ratios rather than center-of-mass energy.

The complementarity of the 240 GeV and 365 GeV energy points is evident: while 240 GeV allows for precise Higgs mass and coupling measurements via $HZ$ production, the 365 GeV stage enhances sensitivity to VBF processes and new physics effects. Therefore, a multi-energy running strategy at FCC-ee maximizes the physics potential for Higgs studies.

\section{Backgrounds}

It is crucial to understand and suppress the Standard Model (SM) background processes that can resemble the 
$ZH$ final state such as $2l+X$ or most probable final state ($2l+2j$). 

The most significant irreducible background comes from processes with the same visible final states as the signal: $e^{+} e^{-} \to ZZ$; this is the dominant background when one $Z$ decays leptonically ($Z\to l^{+} l^{-}$) and the other $Z$ decays hadronically ($Z\to jj$ or $b\bar{b}$), closely resembling $ZH \to l^{+} l^{-} b\bar{b}$ events. Distinguishing the Higgs from the $Z$ requires excellent invariant mass resolution and b-tagging efficiency.
The ISR and BS can lower the effective center-of-mass energy and distort the kinematic distributions, enhancing contributions from processes like $e^+ e^- \to ZZ$ at lower invariant masses. Accurate modeling and correction of ISR effects are necessary for precision studies.

Some reducible backgrounds may be suppressed through selection cuts but still contribute due to detector resolution effects or misidentification: $e^+e^- \to W^+ W^- \to l^+l^- + MET$: although the final state includes missing transverse energy (MET), some $W^+W^-$ events can pass the signal selection if a lepton and two jets are reconstructed similarly to a  $ZH$ candidate.

At $\sqrt{s} = 240$ GeV center-of-mass energy, 
there is no contribution from real top-antitop ($t\bar t$) production to the background. Only off-shell production contributes through $e^+ e^- \to W^+ W^- b\bar b$ ($2\to 4$) process. However the $ZH$ analyses at 240 GeV are not much impacted by this $t^*\bar t^*$ process.
This is also one of the reasons why FCC-ee chooses this energy point for precision Higgs studies. At $\sqrt{s} = 365$ GeV, $t\bar{t}$ is an important and potentially irreducible background for certain $ZH$ final states. Careful analysis is needed to control and subtract it, especially when studying $H \to b\bar{b}$ or other hadronic Higgs decays.

For FCC-ee at $240$ GeV and $365$ GeV, the overlay of multiple $\gamma\gamma \to \text{hadrons}$ events per bunch crossing is smaller but still relevant for precision jet measurements. This background mainly affects jet reconstruction and event pile-up-like contamination, especially in the forward region.
It contributes additional low-$p_T$ tracks and calorimeter clusters, which degrades b-tagging performance, jet energy resolution and recoil mass resolution. After cleaning jets the resolution can be nearly recovered at the mentioned energy scales. Here, it does not dominate our event counts, a quantitative impact at most $10\%$ on mass resolution, however it is important for detector performance and precision observables (especially for hadronic final states), see Ref. \cite{Thomson2016, Heinemeyer2021}.

Here, we are interested in main backgrounds resulting in two opposite sign leptons (electron or muon) and at least two jets (at least one is b-tagged). There may be other relevant backgrounds, but we need to avoid them overlapping with the main $ZZ$ or $t \bar t$ background.

The cross sections for the backgrounds calculated by \texttt{MadGraph5\_aMC@NLO} \cite{Alwall2011} including (ISR+BS) at two different center of mass energies have been shown in Table \ref{tab:backg_cross_sections_ISR}.

\begin{table}[ht]
    \begin{center}
    \caption{Cross sections for the background processes at $\sqrt{s} = 240$ GeV and $\sqrt{s} = 365$ GeV. Respective branching ratios can be used $BR(W \to l{\nu}) = 1.08\times 10^{-1}$ and $BR(Z \to l^+l^-) = 3.37 \times 10^{-2}$ for the final state. \label{tab:backg_cross_sections_ISR}}
    \vspace{0.5em}
    \tiny
\begin{tabular}{|l|c|c|c|c|}
\hline
Process & \multicolumn{2}{c|}{$\sqrt{s} = 240$ GeV} & \multicolumn{2}{c|}{$\sqrt{s} = 365$ GeV} \\  \cline{2-5}
 & $\sigma$ (pb) & $\sigma$ (ISR+BS) (pb) & $\sigma$ (pb) & $\sigma$ (ISR+BS) (pb) \\
\hline
$e^+ e^- \to ZZ$ & $1.162\times 10^0$ & $1.112\times 10^0$ & $6.425 \times 10^{-1}$ & $6.717 \times 10^{-1}$ \\
$e^+ e^- \to W^+ W^-$ & $1.715\times 10^1$ & $1.679\times 10^1$ & $1.078\times 10^1$ & $1.107\times 10^1$ \\
$e^+ e^- \to ZZ\, b\bar{b}$ & $3.293 \times 10^{-4}$ & $2.776 \times 10^{-4}$ & $3.093 \times 10^{-4}$ & $2.699 \times 10^{-4}$ \\
$e^+ e^- \to W^+ W^-\, b\bar{b}$ & $9.638 \times 10^{-5}$ & $8.510 \times 10^{-5}$ & $4.062 \times 10^{-1}$ & $2.616 \times 10^{-1}$ \\
$e^+ e^- \to t\bar{t}$ & -- & -- & $4.883 \times 10^{-1}$ & $3.708 \times 10^{-1}$ \\
$e^+ e^- \to l^+ l^-$ & $6.721\times 10^2$ & $6.758\times 10^2$ & $2.913\times 10^2$ & $2.918\times 10^2$ \\
\hline
\end{tabular}
\end{center}
\end{table}

\section{Detector and event simulation}

The detector design and event simulation strategy play a crucial role in studying the associated production of the Higgs boson with a $Z$ boson at future lepton colliders. In this work, we consider the IDEA (Innovative Detector for Electron–positron Accelerator) detector concept proposed for the FCC-ee, which offers excellent tracking and calorimetric performance, optimized for precision measurements in clean environments.

The IDEA detector consists of: (a) a low-material, high-granularity cylindrical tracker to minimize multiple scattering, (b) a high-resolution vertex detector for precise reconstruction of displaced vertices and impact parameters, (c) a finely segmented calorimeter system for accurate energy measurements of electromagnetic and hadronic showers.

These features collectively enable excellent momentum and energy resolution for charged leptons and jets, which is critical for processes like $e^+e^- \rightarrow ZH$, especially when identifying final-state leptons and reconstructing recoil mass distributions.
Simulated signal and background events were generated using \texttt{MadGraph5\_aMC@NLO} \cite{Alwall2011}, showered with \texttt{PYTHIA8} \cite{Bierlich2022}, and processed through \texttt{Delphes} \cite{Ovyn2009} using the \texttt{IDEA} card. In general electron ID and muon ID efficiencies can be slightly different as they are identified by different sub detectors, however for momentum $p_T > 10$ GeV, here it is assumed to be $98-99 \%$. We combine electron and muon channels as leptons for a phenomenological analysis.

\section{Analysis and results}

The analysis focuses on the production $e^+ e^- \to HZ$, and decay $Z \to l^+ l^-$, where $l = e$ or $\mu$. Event selection criteria are as follows:
\begin{itemize}
\item Exactly the same flavor and two oppositely-charged leptons ($e^\pm$ or $\mu^\pm$) with transverse momentum $p_T > 20$ GeV and pseudorapidity $|\eta| < 2.5$, 
\item Dilepton transverse momentum $p_T^{ll} > 40$ GeV, 
\item Absolute difference from the $Z$ boson mass $|m_{ll} - 91.18~\mathrm{GeV}| < 20$ GeV,
\item Recoil mass for the  system within $120~\mathrm{GeV} < m_{ll} < 140~\mathrm{GeV}$.
\end{itemize}

The efficiency after the $p_T$ cut and $\eta$ cut mentioned in the text is about $65\%$ for the signal. Most of the dilepton events lie in the invariant mass inside the $Z$ boson mass window for signal.

In our analysis, recoil mass distributions were obtained from simulated $ e^+ e^- \rightarrow ZH $ events at both 240 GeV and 365 GeV center-of-mass energies at an integrated luminosity of 5 ab$^{-1}$. The signal manifests as a clear peak around the nominal Higgs mass of 125 GeV \footnote{We use natural units with $\hbar =1$ and $c=1$}. The background contributions, mainly from $ ZZ $ and $ WW $ processes, form a smooth distribution under the peak.
These results demonstrate that the recoil mass method at future circular electron-positron colliders offers a powerful and precise tool for Higgs mass measurement, with complementary benefits at different operating energies.

\begin{figure}[ht]
    \centering
    \includegraphics[width=0.7\textwidth]{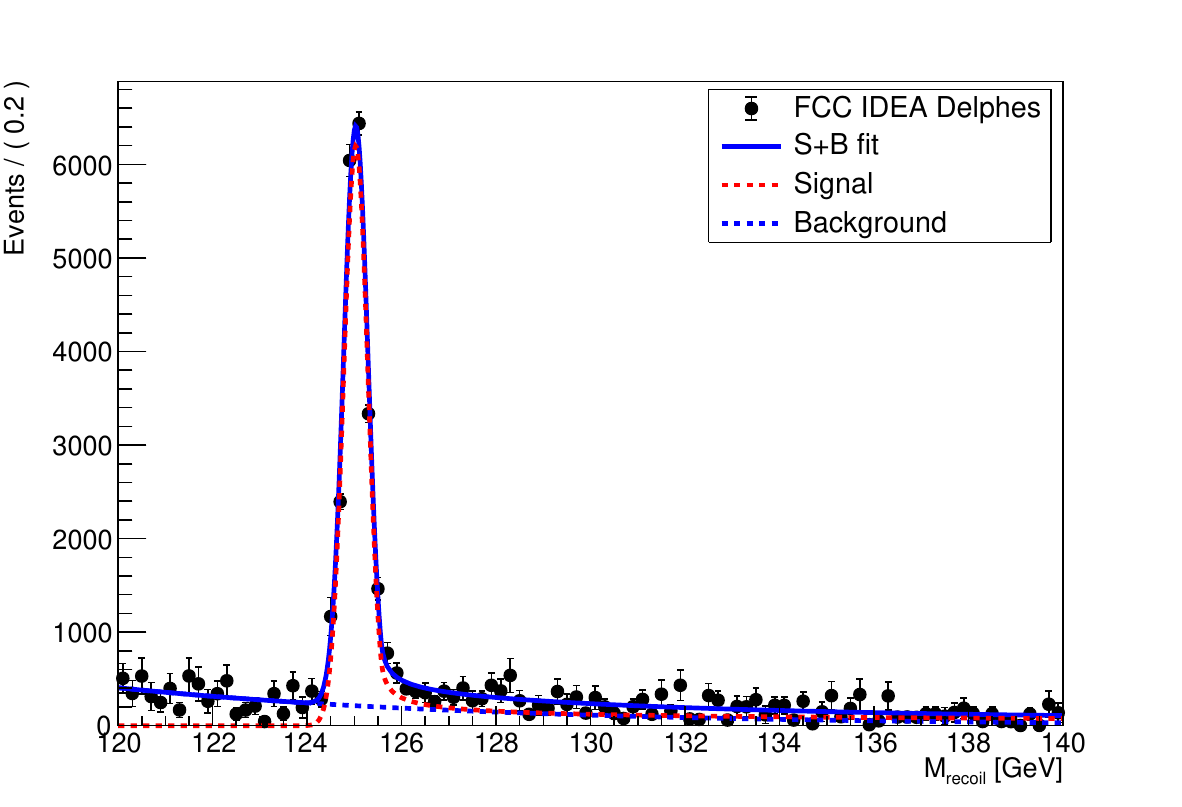}
    \caption{Recoil mass distribution at $\sqrt{s} = 240$ GeV with S+B fit. The events are presented for an integrated luminosity of 5 ab$^{-1}$.}
    \label{fig:recoil240}
\end{figure}

The recoil mass distribution obtained from simulated signal events at $\sqrt{s} = 240$ GeV, with the $Z$ boson decaying leptonically ($Z \rightarrow l^+l^-$) have been shown in Figure~\ref{fig:recoil240}. This distribution was fitted using a signal plus background (S+B) model. 
The fit yields a Higgs boson mass of $
 m_H = 125.032~\text{GeV} \text{ (mean)}$ with $ \sigma_H = 0.256~\text{GeV} \text{ (sigma)}$ from the recoil mass distribution.

This result demonstrates the excellent recoil mass resolution achievable with the IDEA detector at FCC-ee. The mass resolution of approximately 0.25 GeV is consistent with expectations for a clean leptonic final state ($Z \to l^+l^-$) and confirms the ability of the detector and reconstruction methods to precisely resolve the Higgs boson signal. Such precision is crucial for probing the $HZZ$ coupling and for potential deviations from the Standard Model predictions. The distribution exhibits a clear and narrow peak centered around the nominal Higgs mass. This narrow peak around 125 GeV enables accurate extraction of the inclusive $e^+e^- \to HZ$ cross section and related coupling parameters. The cross section is expected to about $0.5\%$ precision at FCC-ee \cite{Azzurri2022}.

The signal was modeled using a Double Side Crystal Ball (DSCB) function from the RooFit package of the data analysis software framework ROOT \cite{Rene1997}, and the background with a smooth polynomial shape. A combined signal-plus-background (S+B) fit was applied to extract the Higgs mass and resolution parameters. The sharp peak near 125 GeV demonstrates the excellent mass resolution achievable with the recoil technique at this energy.

Background contributions, mainly from $ZZ$ and $WW$ processes, form a smooth underlying distribution, well-separated from the Higgs signal. The clean experimental conditions at 240 GeV and the high signal-to-background ratio allow for accurate extraction of Higgs properties.

\begin{figure}[htb]
    \centering
    \includegraphics[width=0.7\textwidth]{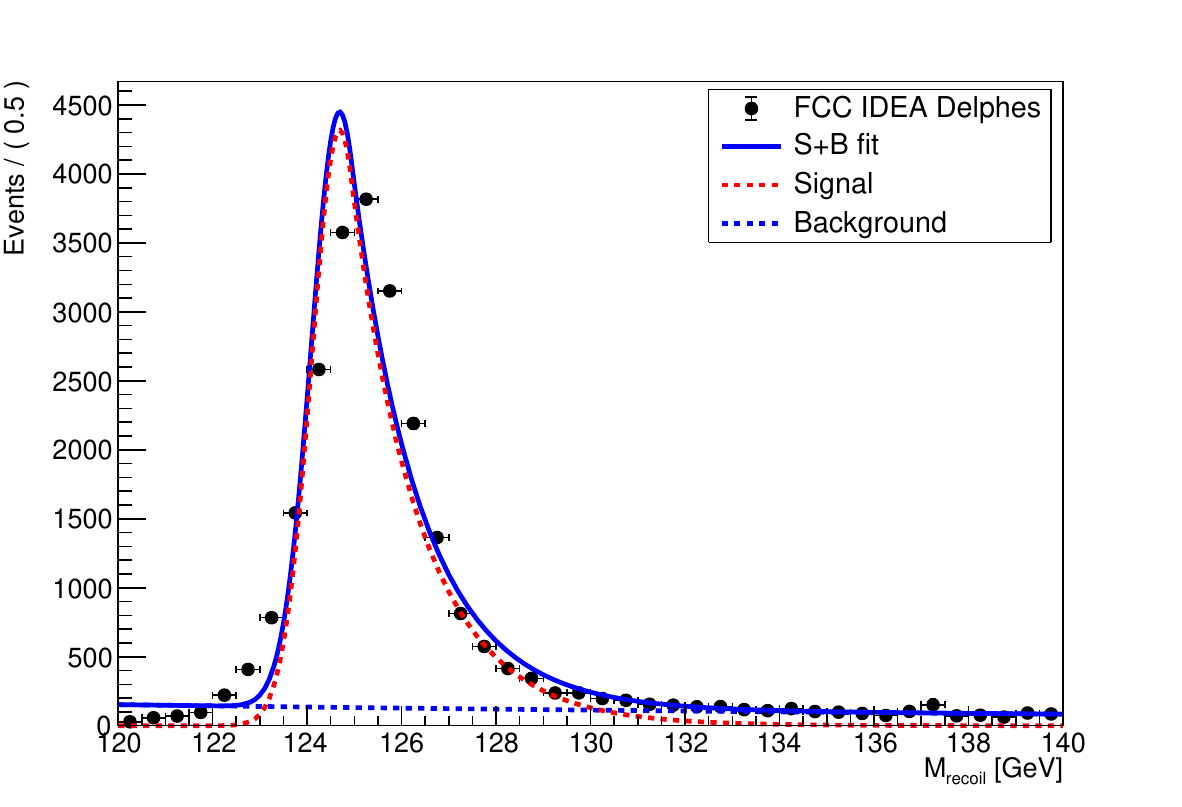}
    \caption{Recoil mass distribution at $\sqrt{s} = 365$ GeV with S+B fit. The events are presented for an integrated luminosity of 5 ab$^{-1}$.}
    \label{fig:recoil365}
\end{figure}

Figure~\ref{fig:recoil365} presents the recoil mass distribution for the process $e^+e^- \rightarrow ZH$ at a center-of-mass energy of $\sqrt{s} = 365$ GeV, where the $Z$ boson decays into a pair of leptons. At this energy, the recoil mass distribution  exhibits a broader peak compared to the 240 GeV case, as expected due to increased beam energy spread and more significant initial/final state radiation effects. The fit to the signal plus background yields: $m_H = 124.69~\text{GeV} \text{ (mean)}$ with $\sigma_H = 0.60~\text{GeV} \text{ (sigma)}$ from the recoil mass distribution.

While the mass resolution at 365 GeV is large compared to the 240 GeV case, the extended kinematic range provides enhanced sensitivity to the high-$p_T$ regime of the final-state leptons. This energy stage is therefore well-suited for complementary studies of the $HZZ$ vertex and possible deviations from the Standard Model expectations in the off-shell or boosted kinematic regions.

The signal peak remains centered near the Higgs boson mass, but appears broader due to the increased effects of initial/final state radiation, beam energy spread, and reduced phase-space constraints and a wider width of the recoil peak indicates a degradation in mass resolution compared to the 240 GeV case. This is expected, as higher center-of-mass energy leads to larger ISR and more pronounced beamstrahlung effects, both of which smear the reconstructed mass distribution.

Despite the increased background activity and broader peak, the signal remains clearly distinguishable. The 365 GeV energy point offers complementary advantages, such as access to higher recoil momentum and sensitivity to potential deviations in differential distributions. These features are valuable for probing anomalous couplings and studying the momentum dependence of the $HZZ$ interaction.

Overall, the recoil mass method remains effective at higher energies, and combining measurements at both 240 GeV and 365 GeV enhances the global sensitivity to Higgs properties and new physics scenarios.

As the result, fitting the recoil mass peak with an appropriate function (DSCB) allows extraction of the Higgs mass and its resolution. As expected, the recoil mass resolution is better at 240 GeV due to smaller BS effects and more constrained kinematics, resulting in a narrower peak and thus a more precise Higgs mass measurement. At 365 GeV, although the recoil mass peak is broader due to increased ISR and beam energy spread, the higher cross section for some background processes requires more detailed background modeling (beam induced pairs, $\gamma\gamma\to$ hadrons, ISR+BS effects, boosted objects (H,Z), lepton momentum/energy scale, etc.).
The signal peak near 125 GeV is clearly visible. The background is modeled by the fitted functions. The 365 GeV simulation data exhibits broader resolution but higher sensitivity at large transverse momentum.

\section{Conclusion}

In this study, the associated production of the Higgs boson with a $Z$ boson at the FCC-ee was investigated using the recoil mass method, focusing on the leptonic decay channel of the $Z$ boson. The analysis was performed at two center-of-mass energies, $\sqrt{s} = 240$ GeV and $\sqrt{s} = 365$ GeV, using fully simulated signal and background events processed through a fast detector simulation with the IDEA detector model.

The recoil mass method enabled a model-independent reconstruction of the Higgs boson without relying on its decay products. This allowed precise measurements of the Higgs boson mass and the cross section of the $e^+e^- \to ZH$ process, which are directly sensitive to the $HZZ$ coupling.

The results showed that: (i) at $\sqrt{s} = 240$ GeV, the recoil mass resolution is optimal, providing a narrow Higgs peak with reduced background contamination; (ii) at $\sqrt{s} = 365$ GeV, although the resolution is slightly degraded due to increased initial state radiation and beam energy spread, the higher recoil momentum regime improves sensitivity to certain new physics effects.

The optimized event selection and fitting procedure demonstrated that both energy stages of FCC-ee are complementary in probing the Higgs sector. These measurements are crucial for testing the Standard Model predictions and searching for potential deviations that may indicate new physics.

The recoil mass method, supported by a clean experimental environment and high luminosity at FCC-ee, remains one of the most powerful tools for precision Higgs physics and model-independent coupling extraction. Future extensions of this work may include the incorporation of hadronic $Z$ decays, differential cross section analyses, and studies of angular distributions to further constrain the Higgs properties and possible anomalous couplings.

\section*{Acknowledgments}
This work was partially supported by Turkish Energy, Nuclear and Mineral Research Agency (TENMAK) under project number 2025TENMAK (CERN) A5.H3.F2-05.

\end{document}